\documentclass[%
 reprint,
 amsmath,amssymb,
 aps,
]{revtex4-1}

\usepackage{graphicx}
\usepackage{dcolumn}
\usepackage{bm}
\usepackage{bbm}
\usepackage{braket}
\usepackage{color}
\usepackage{here}
\usepackage{mathtools}
\usepackage[normalem]{ulem}

\newcommand{\up}{\uparrow}
\newcommand{\down}{\downarrow}
\newcommand{\s}{\sigma}

\begin{document}

\preprint{APS/123-QED}

\title{Spin current at a magnetic junction as a probe of the Kondo state}
\author{T. Yamamoto$^{1}$, T. Kato$^{1}$, and M. Matsuo$^{2,3,4,5}$}
\affiliation{
${^1}$Institute for Solid State Physics, The University of Tokyo, Kashiwa, Japan\\
${^2}$Kavli Institute for Theoretical Sciences, University of Chinese Academy of Sciences, Beijing, China\\
${^3}$CAS Center for Excellence in Topological Quantum Computation, University of Chinese Academy of Sciences, Beijing 100190, China\\
${^4}$Advanced Science Research Center, Japan Atomic Energy Agency, Tokai, 319-1195, Japan\\
${^5}$RIKEN Center for Emergent Matter Science (CEMS), Wako, Saitama 351-0198, Japan
}

\date{\today}

\begin{abstract}
We investigate the spin Seebeck effect and spin pumping in a junction between a ferromagnetic insulator and a magnetic impurity deposited on a normal metal.
By performing a numerical renormalization group calculation, we show that spin current is enhanced by the Kondo effect.
This spin current is suppressed by an increase in temperature or a magnetic field comparable to the Kondo temperature.
Our results indicate that spin transport can be a direct probe of spin excitation in strongly correlated systems.
\end{abstract}

\pacs{Valid PACS appear here}

\maketitle

{\it Introduction.}---
Spin current at a magnetic junction driven by spin pumping (SP)~\cite{Tserkovnyak2002,Tserkovnyak2005} or the spin Seebeck effect (SSE)~\cite{Uchida2010,Bauer2012} has been studied intensively in the field of spintronics~\cite{Zutic2004}.   
Recently, it has been recognized that these effects can be utilized to detect spin-related properties in nano-scale systems~\cite{Han2018}, such as long-range spin transport due to spin-triplet pairs at a ferromagnetic interface~\cite{Jeon2018} and the antiferromagnetic phase transition of a magnetic thin film~\cite{Qiu2016}.  
It is remarkable that measurements of spin current at a magnetic junction are more sensitive even for such a nano-scale thin film than conventional bulk measurement techniques such as NMR and neutron scattering. 
This implies that SP and SSE as well as nonlocal spin valve measurements~\cite{Garzon2005} will enable more detailed measurements of nano-scale spin systems~\cite{Zutic2004,Han2019} that have been considered to be difficult with the conventional bulk measurement techniques.

In this Letter, we focus on the Kondo effect, which is one of the most significant many-body phenomena in condensed matter physics.
We consider magnetic impurities on a metal surface and examine how SP and SSE detect their spin excitation.
The Kondo effect of transition-metal atoms or molecules containing them on a metal surface has been studied for a long time by scanning tunneling microscopy (STM)~\cite{Madhavan1998,Li1998,Manoharan2000,Nagaoka2002,Wahl2004,Zhao2005,Tsukahara2011,Karan2015,Hiraoka2017,Fernandez2021}.
In these STM experiments, the signatures of the Kondo effect were studied by using the differential conductance that reflects the local density of states of magnetic impurities.
In comparison with STM, the present proposal based on SP and SSE has an advantage in that it can access spin excitation in the Kondo state directly.

\begin{figure}[tb]
\includegraphics[width=0.8\columnwidth]{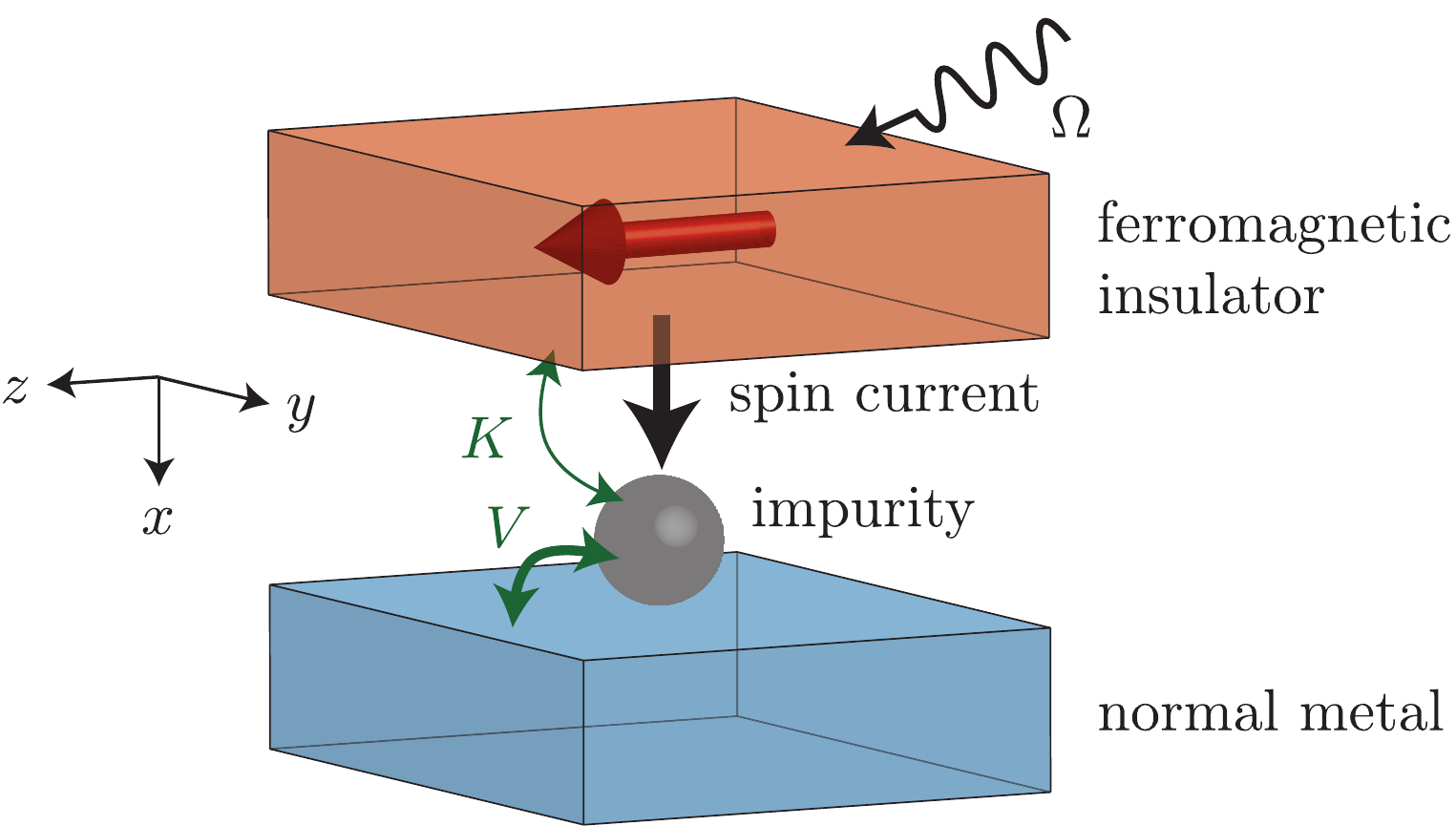}
\caption{Schematic model considered in this Letter.
A magnetic impurity deposited on a normal metal (NM) is weakly coupled with a ferromagnetic insulator (FI).
The spin current from the FI is induced either by a temperature gradient between the NM and the FI (spin Seebeck effect) or an external ac magnetic field applied to the FI (spin pumping).}
\label{fig:setup}
\end{figure}

{\it Model.}---
We consider a junction between a ferromagnetic insulator (FI) and a magnetic impurity deposited on the surface of a normal metal (NM) (see Fig.~\ref{fig:setup}).
The model Hamiltonian consists of three terms, ${\cal H}={\cal H}_{\rm A}+{\cal H}_{\rm FI}+{\cal H}_{\rm I}$.
The magnetic impurity and the NM are described by the impurity Anderson model, whose Hamiltonian is given as
\begin{align}
\label{eq:HA}
{\cal H}_{\rm A}
&=\sum_{\s}\epsilon_dd_{\s}^\dagger d_\s
+Ud_\up^\dagger d_\up d_{\down}^\dagger d_{\down}
-\frac{\hbar\gamma h_{\rm dc}}{2}\left(d^\dagger_\up d_\up-d^\dagger_\down d_\down\right) \nonumber\\
&\quad+\sum_{{\bm k}\s}\epsilon_{\bm k}c_{{\bm k}\s}^\dagger c_{{\bm k}\s}
+\sum_{{\bm k}\s}\left(V_{\bm k}d_\s^\dagger c_{{\bm k}\s}+{\rm h}.{\rm c}.\right) .
\end{align}
Here, $d_\s$ is the annihilation operator of an electron in the magnetic impurity with spin $\s(=\up,\down)$ and $c_{{\bm k}\s}$ is that of a conduction electron in the NM with wave number ${\bm k}$ and spin $\s$.
The energy level of the magnetic impurity and the on-site Coulomb interaction are denoted by $\epsilon_d$ and $U$, respectively, and the chemical potential of the NM at equilibrium is set to be the origin of the energy.
The effect of the magnetic field is incorporated in the Zeeman energy for the magnetic impurity with gyromagnetic ratio $\gamma$ and static magnetic field $h_{\rm dc}$.
The magnetic impurity is hybridized with a conduction band with an energy dispersion $\epsilon_{{\bm k}\s}$ through the coupling constant $V_{\bm k}$.
Here, it is convenient to introduce the hybridization function $\Gamma\equiv\pi\sum_{\bm k}|V_{\bm k}|^2\delta(\hbar\omega-\epsilon_{\bm k})$.
Assuming a ${\bm k}$-independent hybridization ($|V_{\bm k}|\equiv V$) and the wide-band limit, the hybridization function becomes constant, $\Gamma=\pi V^2/(2W)$, where $W$ is the conduction bandwidth.
In this letter, we neglect the spin-orbit coupling (SOC) in the NM as it hardly affect the Kondo effect~\cite{Zitko2011}.

The FI is modeled by localized spins with a Heisenberg-type exchange interaction:
\begin{align}
\label{eq:HFI}
{\cal H}_{\rm FI}=J\sum_{\braket{i,j}}{\bm S}_i\cdot{\bm S}_j-\hbar\gamma h_{\rm dc}\sum_{i}S_i^z,
\end{align}
where ${\bm S}_i$ is a localized spin operator at site $i$, $J(<0)$ is the ferromagnetic exchange coupling constant, and $\braket{i,j}$ indicates a pair of nearest-neighbor sites.

The exchange interaction between the magnetic impurity and the FI is described by
\begin{align}
\label{eq:HI}
{\cal H}_{\rm I}=\sum_{\bm k}\left(K_{\bm k}S^{+}_{\bm k}s_d^{-}+{\rm h}.{\rm c}.\right),
\end{align}
where $s_d^{\pm}=s_d^x\pm is_d^y$ is the spin ladder operator of the magnetic impurity and $S^{\pm}_{\bm k}$ is the Fourier transformation of the spin ladder operator $S_j^{\pm}=S_j^x\pm iS_j^y$.
For simplicity, we assume that the exchange interaction is independent of ${\bm k}$, i.e., $K_{\bm k}\equiv K$. 
Here, we dropped the $z$-component term, $K\sum_{\bm k}S_{\bm k}^zs_d^z$, which describes an exchange field, i.e., an effective magnetic field added in the Anderson-type Hamiltonian~\eqref{eq:HA} after replacing $S^z(\bm r={\bm 0})$ with an averaged one $\braket{S^z}$.
This approximation, that neglects the proximity effect from the FI to the magnetic impurity~\cite{Zutic2019}, is justified when the exchange coupling is sufficient weak, $K\ll k_{\rm B}T_{\rm K}$.
We note that this exchange field may affect spin current in general when $K \gtrsim k_{\rm B}T_{\rm K}$.

{\it Spin-wave approximation.}---
Here, we employ the spin-wave approximation (SWA) based on the Holstein-Primakoff transformation~\cite{Holstein1940}.
Assuming that the spin magnitude $S_0$ is sufficiently large and that the temperature is much lower than the transition temperature, the Hamiltonian of the FI is approximated as non-interacting magnons,
\begin{align}
\label{eq:Hmag}
{\cal H}_{\rm FI}
\approx\sum_{\bm k}\hbar\omega_{\bm k}b_{\bm k}^\dagger b_{\bm k}, 
\end{align}
where $b_{\bm k}$ is a bosonic annihilation operator.
The magnon dispersion is described as $\hbar\omega_{\bm k}=Dk^2+E_0$ with spin stiffness $D=|J|S_0a^2$ ($a$ is the lattice constant) and Zeeman energy $E_0=\hbar\gamma h_{\rm dc}$.

{\it Spin current.}---
The operator for spin current flowing from the FI is defined as $I_s\equiv\hbar \dot{s}_{\rm tot}^z$, where $s_{\rm tot}^z$ is the $z$ component of the total spin of the conduction electrons in the NM~\cite{Matsuo2018,Kato2019,Kato2020}.
Using the Keldysh formalism, the spin current is calculated within second-order perturbation with respect to the interaction $K$ as~\cite{Matsuo2018,Kato2019,Kato2020,Rammer1986,Bruus}
\begin{align}
\Braket{I_s}
=&\frac{2\hbar K^2}{\pi}\int_{-\infty}^\infty d\omega~
{\rm Im}\left[\chi^\mathrm{R}(\omega)\right]{\rm Im}\left[G^\mathrm{R}(\omega)\right] \nonumber\\
\label{eq:Is}
&\times\left[n_{\rm FI}(\omega) -n_{\rm NM}(\omega)\right]. 
\end{align}
Here, $\chi^\mathrm{R}(\omega)$ and $G^\mathrm{R}(\omega)$ are the Fourier components of the retarded spin correlation functions of the magnetic impurity and the FI, defined as
\begin{align}
\chi^\mathrm{R}(t)&=\frac{1}{i\hbar}\theta(t)\Braket{\left[s_d^+(t),s_d^-(0)\right]}_{\rm A}, \\
G^\mathrm{R}(t)&=\sum_{\bm k}\frac{1}{i\hbar}\theta(t)\Braket{\left[S_{\bm k}^+(t),S_{\bm k}^-(0)\right]}_{\rm FI},
\end{align}
where $\braket{\cdots}_{{\rm A}/{\rm FI}}$ denotes the thermal average with respect to ${\cal H}_{{\rm A}/{\rm FI}}$.
The nonequilibrium distribution function of the FI is defined with the lesser component of the spin correlation function, $G^<(\omega)$, as $n_{\rm FI}(\omega)=G^<(\omega)/(2i{\rm Im}[G^\mathrm{R}(\omega)])$, while $n_{\rm NM}(\omega)$ is the Bose-Einstein distribution function of the NM with temperature $T_{\rm NM}$.

{\it Spin Seebeck effect.}---
Let us first consider the case that the FI is in equilibrium in the absence of a dc magnetic field ($h_{\rm dc}=0$).
In this situation, $n_\mathrm{FI}(\omega)$ becomes the Bose-Einstein distribution with the temperature $T_\mathrm{FI}$.
Now, let us suppose that the temperature difference between the NM and the FI, $\delta T=T_{\rm FI}-T_{\rm NM}$, is small enough that the spin current can be expanded with respect to it as $\Braket{I_s}\approx G_s\delta T+{\cal O}(\delta T^2)$, where $G_s$ is the linear spin conductance,~\cite{Matsuo2018,Kato2019}
\begin{align}
G_s
=G_{s,0}\int_0^{\infty}d(\hbar\omega)~
\frac{{\rm Im}\left[\chi^\mathrm{R}(\omega)\right]}{\sqrt{\hbar\omega/E_{\rm c}}}
\left[\frac{(\hbar\omega/2k_{\rm B}T)}{\sinh(\hbar\omega/2k_{\rm B}T)}\right]^2.
\end{align}
Here, $G_{s,0}=6k_{\rm B}N_{\rm FI}S_0(K/E_{\rm c})^2$, $E_{\rm c}=(6\pi^2)^{2/3}D/a^{2}$ is the cutoff energy, and $T$ is the average temperature of the NM and the FI.

{\it Spin pumping.}---
To consider SP, we take a weak ac magnetic field, $h_{\rm ac}$, into account by an additional Hamiltonian,
\begin{align}
{\cal V}=-\frac{\hbar\gamma h_{\rm ac}}{2}\sqrt{N_{\rm FI}}\left(e^{-i\Omega t}S_{\bm 0}^-+e^{i\Omega t}S_{\bm 0}^+\right),
\end{align}
where $\Omega$ is a microwave frequency~\footnote{We neglected spin pumping on the magnetic impurity induced by the ac magnetic filed.
We note that this effect does not matter in the Kondo regime because the local moment on the magnetic impurity is screened.}.
The nonequilibrium distribution of the FI, $n_{\rm FI}$, can be evaluated from the lesser and retarded components of the correlation function within second-order perturbation.
Accordingly, we obtain the analytical formula of the spin current induced by the ac magnetic field as~\cite{Matsuo2018,Kato2019}
\begin{align}
\Braket{I_s}
&=I_{s,0}\frac{E_{\rm c}^3~{\rm Im}\left[\chi^\mathrm{R}(\Omega)\right]}{(\hbar\Omega-E_0)^2+(\alpha\hbar\Omega)^2},
\label{eq:formulaSP}
\end{align}
where $I_{s,0}=2N_{\rm FI}(KS_0\hbar\gamma h_{\rm ac})^2/E_{\rm c}^3$ and $\alpha$ is a phenomenological parameter describing the Gilbert damping~\footnote{The spin relaxation in a bulk FI is caused by the magnon-magnon and magnon-phonon scatterings and is usually incorporated phenomenologically in the form of the Gilbert damping.
In the same way, we employ the phenomenological Gilbert damping to take in the spin relaxation in the FI.}.
Here, we have assumed that the temperatures of the NM and the FI are the same, i.e., $T=T_{\rm NM}=T_{\rm FI}$.
We stress that this setup indeed enables us to access the imaginary part of the dynamic spin susceptibility, ${\rm Im} [\chi^\mathrm{R}(\omega)]$,~\cite{Yamada1975_53,Yamada1975_54,Bickers1987,Hanl2014} directly by measuring the spin current (see Eq.~(\ref{eq:formulaSP})).
In the case of SP, in order to stabilize the magnetization of the FI against the external dc magnetic field, we introduce a uniaxial magnetic anisotropy by adding a term ${\cal H}_{\rm aniso}=-\hbar d/(2N_{\rm FI}S_0)\sum_{i}(S_i^{z})^2$ to the Hamiltonian of the FI.
It changes the magnon dispersion to $E_0=\hbar\gamma h_{\rm dc}+\hbar d$.
For simplicity, we assume that the uniaxial magnetic anisotropy is so strong that the dc magnetic-field effect is negligible, i.e., $E_0\simeq\hbar d$.

{\it Numerical renormalization group method.}---
To evaluate the spin current, we need to calculate the imaginary part of the retarded spin correlation function of the magnetic impurity, ${\rm Im}[\chi^\mathrm{R}(\omega)]$.
To do so, we employed the numerical renormalization group (NRG) method~\cite{Wilson1975,Bulla2008} using the reduced density matrix approach associated with the complete Fock space basis~\cite{Anders2005,Anders2006,Peters2006}.
In addition, to obtain smooth curves for the dynamical correlation function, we use the broadening kernel that interpolates between two common broadening functions: Gaussian and logarithmic Gaussian (see~\cite{Weichselbaum2007} for more details).
For the NRG data presented in this Letter, we chose typical values of the logarithmic discretization parameter, number of kept states, and broadening width parameter, i.e., $\Lambda=2.0$, $N_{\rm kept}=1024$, and $b=0.7$, respectively.
Moreover, the numerical results were calculated at large $u=U/(\pi\Gamma)\ge 4$ for which the Kondo singlet is well developed in small magnetic fields and at low temperatures.
The Kondo temperature $T_{\rm K}$ was determined numerically from the static spin susceptibility at zero temperature via $\chi_0(T=0)=1/(4T_{\rm K})$.

\begin{figure}[tb]
\includegraphics[width=0.9\columnwidth]{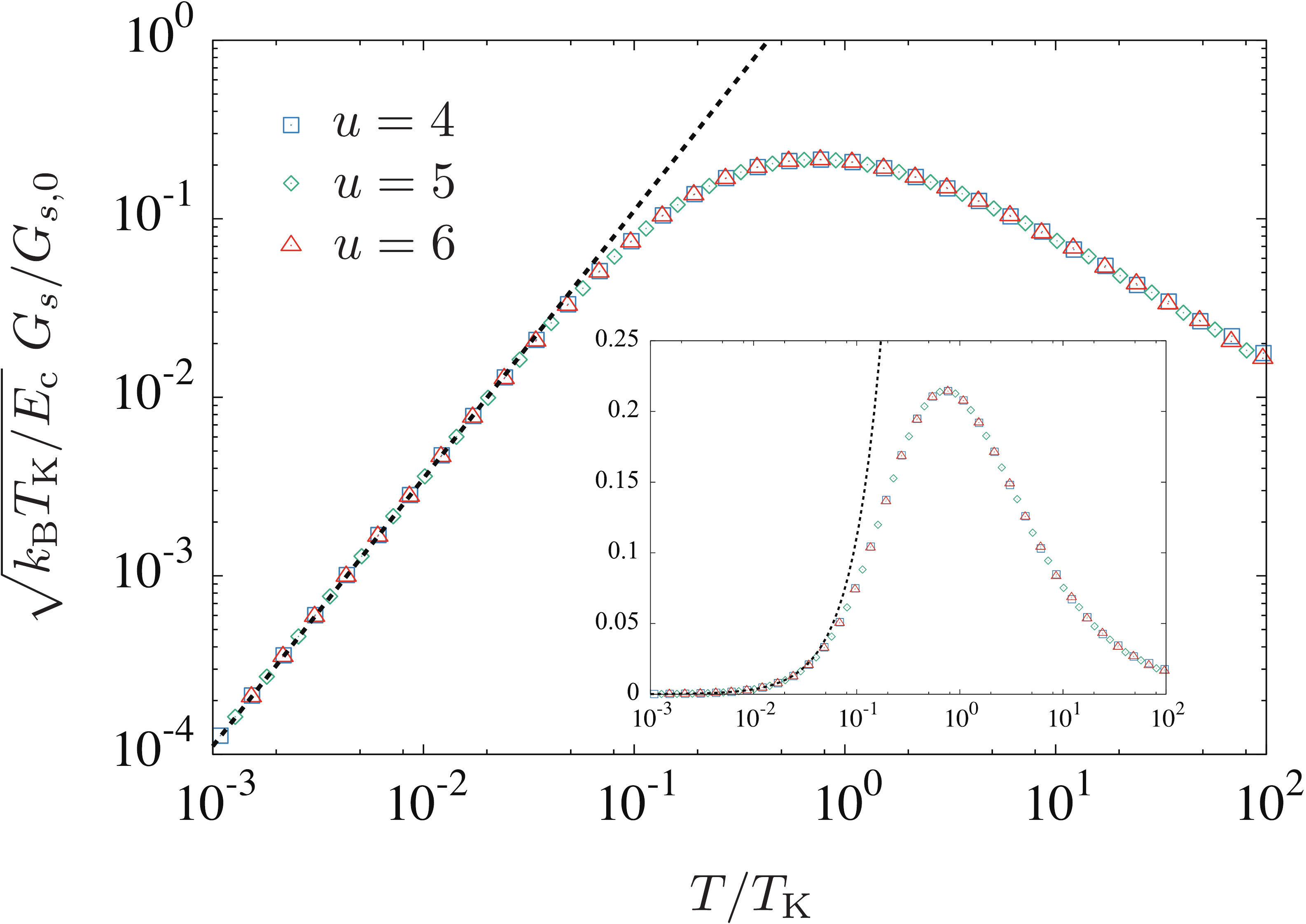}
\caption{Temperature dependence of linear spin conductance calculated by the NRG method for $u=U/(\pi\Gamma)=4,5$, and $6$ with $\Gamma/W=0.005$, $\epsilon_d=-U/2$, $\Lambda=2.0$, $N_{\rm kept}=1024$, and $b=0.7$.
The dotted line represents the low-temperature asymptote given by Eq.~\eqref{eq:SSE_Is_Korringa}.
The inset shows a normal scaled view for the $y$-axis of the main panel.}
\label{fig:SSE_Is_vs_T}
\end{figure}

{\it Results for spin Seebeck effect.}---
The spin current induced by the spin Seebeck effect is shown in Fig.~\ref{fig:SSE_Is_vs_T}.
All the results for $u=4,5,$ and $6$ fall on one universal curve if the temperature and spin conductance are scaled by $T_{\rm K}$ and $(k_{\rm B}T_{\rm K})^{-1/2}$, respectively.
This scaling is an indication of the Kondo effect.
The spin conductance has a peak near the Kondo temperature and shows a power-law behavior with exponent $3/2$ at low temperatures.
The low-temperature power-law behavior is explained by the Korringa relation~\cite{Shiba1975} for the imaginary part of the low-frequency spin susceptibility.
In fact, it gives as asymptotically exact expression for the spin conductance,
\begin{align}
\label{eq:SSE_Is_Korringa}
\sqrt{\frac{k_{\rm B}T_{\rm K}}{E_{\rm c}}}\frac{G_s}{G_{s,0}}
\sim\frac{\pi c}{\sqrt{2}}~\left(\frac{T}{T_{\rm K}}\right)^{3/2},
\end{align}
where $c=\int_0^\infty dx x^{5/2}/\sinh^2x\approx1.58$.
Our numerical calculation agrees with this asymptotic expression for $T\ll T_{\rm K}$, as shown in Fig.~\ref{fig:SSE_Is_vs_T}.

\begin{figure}[tb]
\includegraphics[width=0.9\columnwidth]{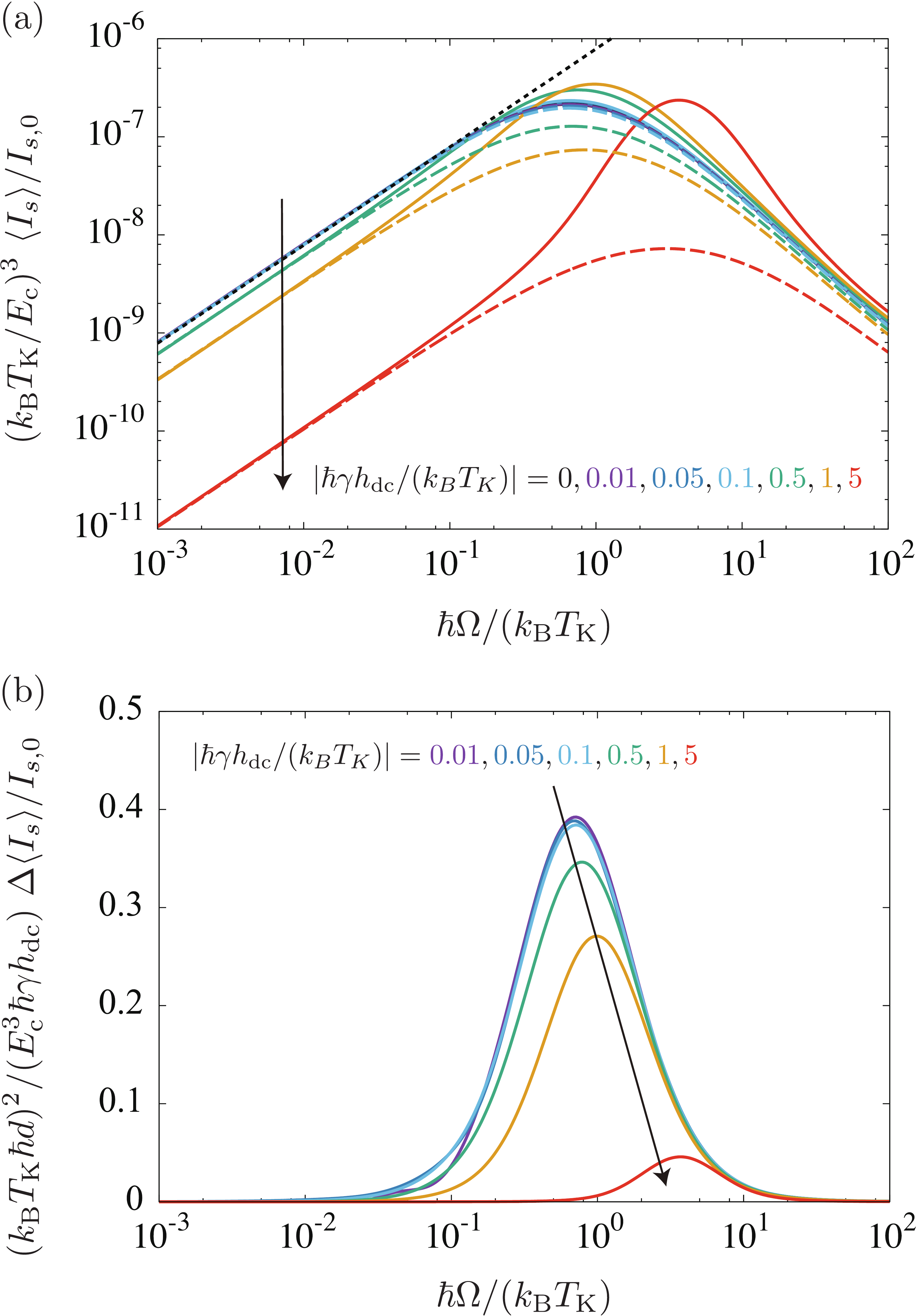}
\caption{(a) Spin current as a function of the resonant frequency $\Omega$ evaluated with the NRG calculation for $u=U/(\pi\Gamma)=4$, $\Gamma/W=0.005$, $\epsilon_d=-U/2$, $\hbar d/k_{\rm B}T_{\rm K}=10^3$, $\alpha=0.1$, and different magnetic fields, $h_{\rm dc}>0$ (solid lines) and $h_{\rm dc}<0$ (dashed lines), at zero temperature.
The dotted line represents the low-frequency asymptote using the Korringa relation.
(b) The difference in spin current $\Delta \braket{I_s}$ between $h_{\rm dc}>0$ and $h_{\rm dc}<0$.
The NRG parameters are as in Fig.~\ref{fig:SSE_Is_vs_T}.}
\label{fig:pumping_Is_vs_omega}
\end{figure}

\begin{figure}[tb]
\includegraphics[width=0.9\columnwidth]{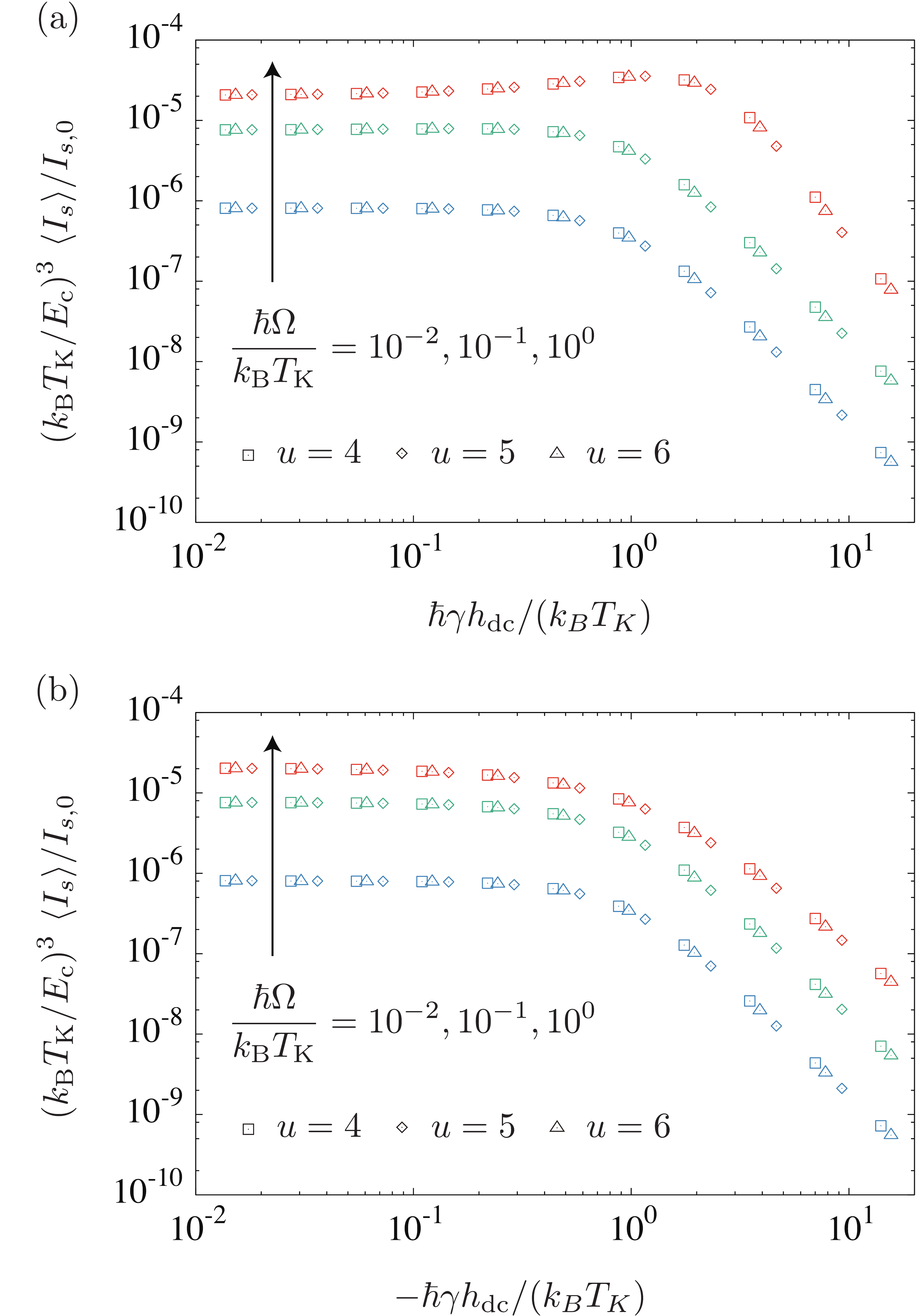}
\caption{DC magnetic-field dependence of the spin current evaluated with the NRG method for different $u=4,5$, and $6$ and $\hbar\Omega/(k_{\rm B}T_{\rm K})=10^{-2}$ (blue), $10^{-1}$ (green), and $10^{0}$ (red) for (a) $h_{\rm dc}>0$ and (b) $h_{\rm dc}<0$.
The other parameters are as in Fig.~\ref{fig:SSE_Is_vs_T}.}
\label{fig:pumping_Is_vs_h}
\end{figure}

{\it Results for spin pumping.}---
Next, let us examine the spin current induced by SP.
Fig.~\ref{fig:pumping_Is_vs_omega}~(a) shows the spin current for $u=4$ as a function of the resonant frequency $\Omega$ when $|\hbar \gamma h_{\rm dc}/k_{\rm B}T_{\rm K}|=0,0.01,0.05,0.1,0.5,1,$ and $5$.
The solid and dashed curves correspond to the cases that the dc magnetic field is in the same direction ($h_{\rm dc}>0$) and in the opposite direction ($h_{\rm dc}<0$) to the magnetization of the FI, respectively.
Note that the magnetization of the FI is fixed in the $+z$ direction because of the uniaxial magnetic anisotropy.
At small magnetic fields compared with the Kondo temperature, $|\hbar\gamma h_{\rm dc}|\ll k_{\rm B}T_{\rm K}$, the spin current has a peak at $\hbar\Omega\simeq k_{\rm B}T_{\rm K}$ and linearly depends on the resonant frequency at low frequencies, as expected from the Korringa relation~\cite{Shiba1975}.
When the dc magnetic field becomes comparable to the Kondo temperature scale, $|\hbar\gamma h_{\rm dc}|\gtrsim k_{\rm B}T_{\rm K}$, the peak of the spin current shifts to a high frequency corresponding to the magnetic field.
At the same time, the spin current deviates rapidly from the Korringa relation and is suppressed strongly, as shown in Fig.~\ref{fig:pumping_Is_vs_omega}, because the Kondo singlet is broken by the magnetic field.
This strong suppression of spin current is a clear indication of the Kondo effect in the SP measurement.
Comparing the spin currents for $h_{\rm dc}>0$ and $h_{\rm dc}<0$ reveals the asymmetric property for reversal of the dc magnetic field with respect to the magnetization.
The magnitude of the spin current at the maximum value, $\hbar\Omega\simeq {\rm max}(k_{\rm B}T_{\rm K},\hbar\gamma h_{\rm dc})$, is estimated as $(e/\hbar) I_s \sim 10^4~{\rm Am^{-2}}$ for relevant materials, YIG for the FI, Cu for the NM, and Co adatoms for the magnetic impurity, where the parameters are taken as $T_{\rm K}=92~{\rm K}$~\cite{Wahl2004}, $S_0=16$, $\gamma=1.76\times10^7~{\rm Oe^{-1}{\rm s}^{-1}}$, $h_{\rm ac}=0.11~{\rm Oe}$, $a=1.24~{\rm nm}$, $D=9.02\times10^{-39}~{\rm Jm^{2}}$~\cite{Kajiwara2010}, $N_{\rm FI}\sim10^{14}~{\rm m^{-2}}$, and $K=0.01k_{\rm B}T_{\rm K}$.

Fig.~\ref{fig:pumping_Is_vs_omega}~(b) shows the difference in spin current between positive and negative dc magnetic fields, $\Delta \braket{I_s}=\braket{I_s}_{h_{\rm dc}}-\braket{I_s}_{-h_{\rm dc}}$ as a function $\hbar \Omega/(k_\mathrm{B}T_\mathrm{K})$.
We find that for a weak magnetic field $\gamma h_{\rm dc} \ll k_{\rm B}T_{\rm K}$, the numerical results fall on a universal curve when the horizontal axis is scaled by $\hbar\gamma h_{\rm dc}/(k_{\rm B}T_{\rm K}\hbar d)^2$.
When the dc magnetic field becomes comparable to the Kondo temperature scale, $\Delta \braket{I_s}$ is suppressed and is no longer on the universal curve~\footnote{The difference in spin current $\Delta\braket{I_s}$ is related to the impurity magnetization $m_{\rm imp}$, which is expressed for large $u$ as $m_{\rm imp}=\tilde{h}/2+{\cal O}(\tilde{h}^3)$ for $\tilde{h}\ll1$ and $m_{\rm imp}=1/2+{\cal O}(\log \tilde{h})$ for $\tilde{h}\gg 1$, where $\tilde{h}=\hbar\gamma h_{\rm dc}/(k_{\rm B}T_{\rm K})$ is a scaled magnetic field~\cite{Kondo}.
Therefore, $\Delta\braket{I_s}$ depends on the dc magnetic field linearly for $\tilde{h}\ll1$ and is almost independent of it for $\tilde{h}\gg1$.}.
This behavior of the spin current is also an indication of the Kondo effect.
Note that the asymmetric part of the spin current with respect to the magnetic field is related to the dynamic spin correlation function between the $x$- and $y$-components.

Finally, we show the magnetic-field dependence of the spin current induced by SP at zero temperature for $\hbar\Omega/(k_{\rm B}T_{\rm K})=10^{-2}$, $10^{-1}$, and $10^{0}$ in Fig.~\ref{fig:pumping_Is_vs_h}.
The NRG results for $u=4,5$, and $6$ fall on one universal curve for a fixed $\hbar\Omega/(k_{\rm B}T_{\rm K})$ when the Zeeman energy and spin current are scaled by $k_{\rm B}T_{\rm K}$ and $(k_{\rm B}T_{\rm K})^{-3}$, respectively.
From Fig.~\ref{fig:pumping_Is_vs_h}, it is clear that the spin current is almost constant for small magnetic fields, $|\hbar\gamma h_{\rm dc}|\ll k_{\rm B}T_{\rm K}$, but decays when the Zeeman energy exceeds the Kondo temperature~\footnote{We note that one cannot use the Korringa relation at higher magnetic fields, $|\hbar\gamma h_{\rm dc}|>k_{\rm B}T_{\rm K}$, even for asymptotically low frequency because the dynamic spin susceptibility becomes  anisotropic, ${\rm Im}[\chi_{+-}(\omega)]\ne2{\rm Im}[\chi_{zz}(\omega)]$~\cite{Garst2005}.}.

{\it Effect of direct NM-FI exchange coupling.}---
In general, spin current induced by direct exchange coupling between the NM and the FI should exist in addition to the spin current through the magnetic impurity.
From Eq.~\eqref{eq:Is}, its contribution can be evaluated by replacing the imaginary part of the dynamic spin susceptibility of the magnetic impurity, ${\rm Im}[\chi^\mathrm{R}(\omega)]$, with that of the NM, ${\rm Im}[\chi^{\mathrm{R}}_\mathrm{NM}(\omega)]=\pi\hbar\omega/(2W)^2$.
Since ${\rm Im}[\chi^\mathrm{R}(\omega)]\sim \pi\hbar\omega/(2k_{\rm B}T_{\rm K})^2 \gg {\rm Im}[\chi^{\mathrm{R}}_\mathrm{NM}(\omega)]$, the spin current induced by direct NM/FI coupling is sufficiently small compared with that through the magnetic impurity when $u$ is large enough to satisfy $k_{\rm B}T_{\rm K}\ll W$.
This also indicates that the spin current is largely enhanced by the Kondo effect.
In experiments, this enhanced spin current can be detected by examining dependence of concentration of magnetic impurities or by observing its strong suppression due to the breakdown of the Kondo effect by the magnetic field or the increase of the temperature.

{\it Summary.}---
We considered spin transport in a junction between a ferromagnetic insulator and a magnetic impurity deposited on a normal metal (NM) and investigated the spin current due to either the spin Seebeck effect or spin pumping using numerical renormalization group calculations.
We found Kondo signatures in the spin transport at sufficiently low temperatures, $T\ll T_{\rm K}$, and small magnetic fields, $|\hbar\gamma h_{\rm dc}|\ll k_{\rm B}T_{\rm K}$.
A large magnetic field, which breaks the Kondo singlet, strongly suppresses the spin current.
The setup studied here can also be realized in a quantum dot system coupled to a FI.
We hope that our work will motivate development of a new probe to investigate strongly correlated systems, such as in the Kondo problem, in spin transport.

We would like to thank Rui Sakano for helpful comments.
This work was supported by Grant-in-Aid for JSPS Fellows Grant Number 20J11318 (TY), by JSPS KAKENHI Grant Numbers 20K03831 (TK), and by 20H01863 and 21H04565 (MM).
MM is partially supported by the Priority Program of Chinese Academy of Sciences, under Grant No. XDB28000000.

\bibliography{spintronics}

\end{document}